\documentclass[11pt]{article}
\fontfamily{times}
\usepackage{graphicx}
\usepackage{geometry}
\usepackage{amsmath}
\usepackage{natbib}
\usepackage{graphicx}
\usepackage[font=footnotesize,labelfont=bf,skip=0pt]{caption}
\usepackage{booktabs}

\newcommand{\templatefigures}[1]
{\noindent
\begin{minipage}{2cm}
\begin{center}
  \centering
	\vspace{-1cm}
  \includegraphics[scale=0.2]{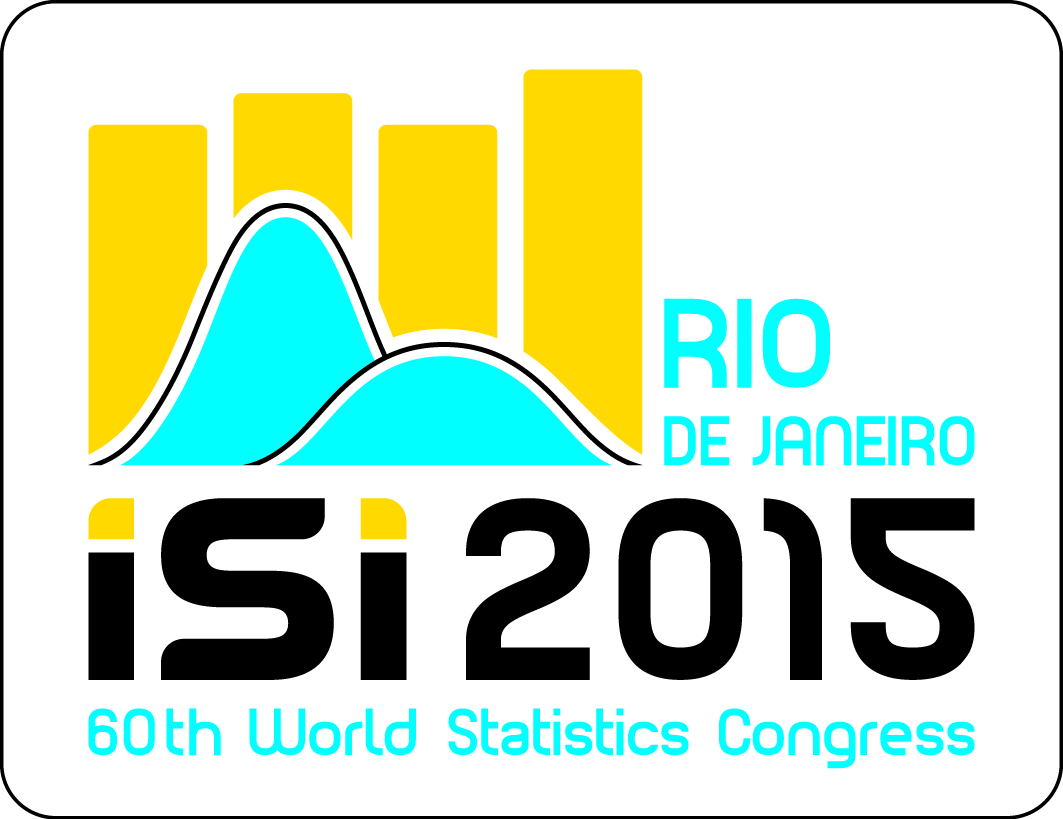}\\
\end{center}
\end{minipage}
\quad
\begin{minipage}{12cm}
\hspace*{6.8cm}
\end{minipage}
\quad
\begin{minipage}{2cm}
\begin{center}
\vspace{-0.9cm}
\includegraphics[scale=0.35]{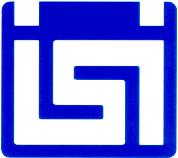}\\
\end{center}

\end{minipage}

\vskip0.2cm
}

\def\OR{{\rm OR}}
\def\bin{{\rm Bin}}
\def\pois{{\rm Pois}}
\def\beta{{\rm Beta}}
\def\cc{{\rm cc}}
\def\obs{{\rm obs}}
\def\half{\hbox{$1\over2$}}

\def\E{{\rm E}} 
\def\Var{{\rm Var}}
\def\prof{{\rm prof}} 
\def\midd{\,|\,}

\def\ml{{\rm ml}} 
\def\mcl{{\rm mcl}} 
 
\def\appr{{\rm approx}} 

\begin{document}
\templatefigures{}

\small{

\begin{center}
\textbf{Optimal inference via confidence distributions for two-by-two tables \\ modelled as Poisson pairs: fixed and random effects}
\end{center}

\begin{center}
{C{\'e}line Cunen*}\\
{University of Oslo, Oslo, Norway - cmlcunen@math.uio.no}\\ 
\vspace{0.5cm}

{Nils Lid Hjort}\\
{University of Oslo, Oslo, Norway - nils@math.uio.no}\\

\end{center}

\begin{center}
{\bf Abstract}
\end{center}

\setlength{\parindent}{0pt}

This paper presents methods for meta-analysis of $2 \times 2$ tables, both with and without allowing heterogeneity in the treatment effects. 
Meta-analysis is common in medical research, but most existing methods are unsuited for $2 \times 2$ tables with rare events.
Usually the tables are modelled as pairs of binomial variables, but we will model them as Poisson pairs.
The methods presented here are based on confidence distributions, and offer
optimal inference for the treatment effect parameter. We also propose an optimal method for inference on the ratio between treatment effects, and illustrate our
methods on a real dataset.\\

}\\

\setlength{\parindent}{0pt}

{\bf 1. Introduction}

Meta-analysis is a class of widely used methods for combination of information from independent studies. In medicine, meta-analyses of $2 \times 2$  tables are especially common. There
are for example many studies investigating whether a certain drug is related to harmful side-effects. The usual set-up is to divide patients into a treatment 
and control group, and count the number of side-effect events in both groups.
If the side-effect in question is very rare, individual studies can have low power, and meta-analysis is therefore 
especially important \citep{liu2014}.
The most common methods for meta-analysis of $2 \times 2$ tables rely on large sample approximations, which makes them unsuited for analysis of tables with rare events \citep{bradburn2007}. 
This applies for example to the well-known Mantel-Haenszel method, the Peto method and the DerSimonian and Laird method (\cite{bradburn2007} and \cite{liu2014}).
Recently \cite{liu2014} introduced certain confidence distributions (CDs) based methods for this type of analysis. In this paper we present different CD-based methodology.
Our methods utilise certain optimality theorems which in particular lead to CDs with a uniformly higher confidence power than for CDs constructed via the methods of \citet{liu2014}. \\
The number of events (cell-counts) in $k$ $2 \times 2$  tables are often modelled as pairs of independent binomial random variables ($Y_{0,i},Y_{1,i}$),
with sample sizes ($m_{0,i},m_{1,i}$) and event probabilities ($p_{0,i},p_{1,i}$). 
Here $Y_{0,i}$ is the number of events among the control group in study $i$ and $Y_{1,i}$ is the number of events in the treatment group in study $i$. 
Usually, the primary interest is not in the event probabilities themselves,
but rather in the potential discrepancy between the event probabilities in the two groups. Whereas such a discrepancy can be defined in different fashions, the usual 
way in biostatistics is via the logistic transform: 
\begin{equation*}
  p_{0,i} = \frac{e^{\theta_i}}{1+e^{\theta_i}}\quad \text{and} \quad p_{1,i} = \frac{e^{\theta_i+\psi_i}}{1+e^{\theta_i+\psi_i}} \text{ .}
\end{equation*}
In particular, the odds ratio is
\begin{equation*}
  \OR_i = \frac{p_{1,i}/(1-p_{1,i})}{p_{0,i}/(1-p_{0,i})} = \frac{e^{\theta_i+\psi_i}}{e^{\theta_i}} = e^{\psi_i} .
\end{equation*}
In the case of small event probabilities, the cell-counts can alternatively be modelled as pairs of independent Poisson variables, 
\begin{equation} \label{eq:5}
  Y_{0,i} \sim \pois(e_{0,i}\lambda_{i}) \quad \text{and} \quad Y_{1,i} \sim \pois(e_{1,i}\lambda_{i}\gamma_i) .
\end{equation}
The argument for this is that binomial variables converge in distribution to Poisson when the sample size is large and the event probabilities are small. 
Here $e_{0,i}$ and $e_{1,i}$ are known exposure
weights reflecting the sample sizes, i.e.~we will expect more cases from a larger study,
than a smaller one, independent of the event probabilities. We choose
\begin{equation*}
  e_{0,i} = m_{0,i}/100  \quad \text{and} \quad e_{1,i} = m_{1,i}/100 . 
\end{equation*}
Thus, $\lambda_{i}$ can be interpreted as the risk of event per 100 patients in the control group, and $\gamma_i$ as the multiplicative parameter associated with the potentially 
different risk (per 100 patients) for patients in the treatment group compared to the 
control group (henceforth called the treatment effect).
Inference concerning odds ratio in the binomial model and $\gamma_i$ in the Poisson model is similar when event probabilities are small and samples sizes are large. In fact, 
\begin{equation*}
  \OR_i = \frac{p_{1,i}/(1-p_{1,i})}{p_{0,i}/(1-p_{0,i})} \approx  \frac{p_{1,i}}{p_{0,i}} \approx \frac{e_{1,i}\lambda_{i}\gamma_i/m_{1,i}}{e_{0,i}\lambda_{i}/m_{0,i}} 
  = \frac{\lambda_{i}\gamma_i}{\lambda_{i}} = \gamma_i .
\end{equation*}
The $\lambda_i$ parameters vary from table to table. In our paper we shall consider the fixed effect model, 
where the $\gamma_i$ are equal to a common $\gamma$, as well as random effect models, where the $\gamma_i$ are allowed to differ.

\begin{figure}[h] 
\begin{center}
\includegraphics[scale=0.5]{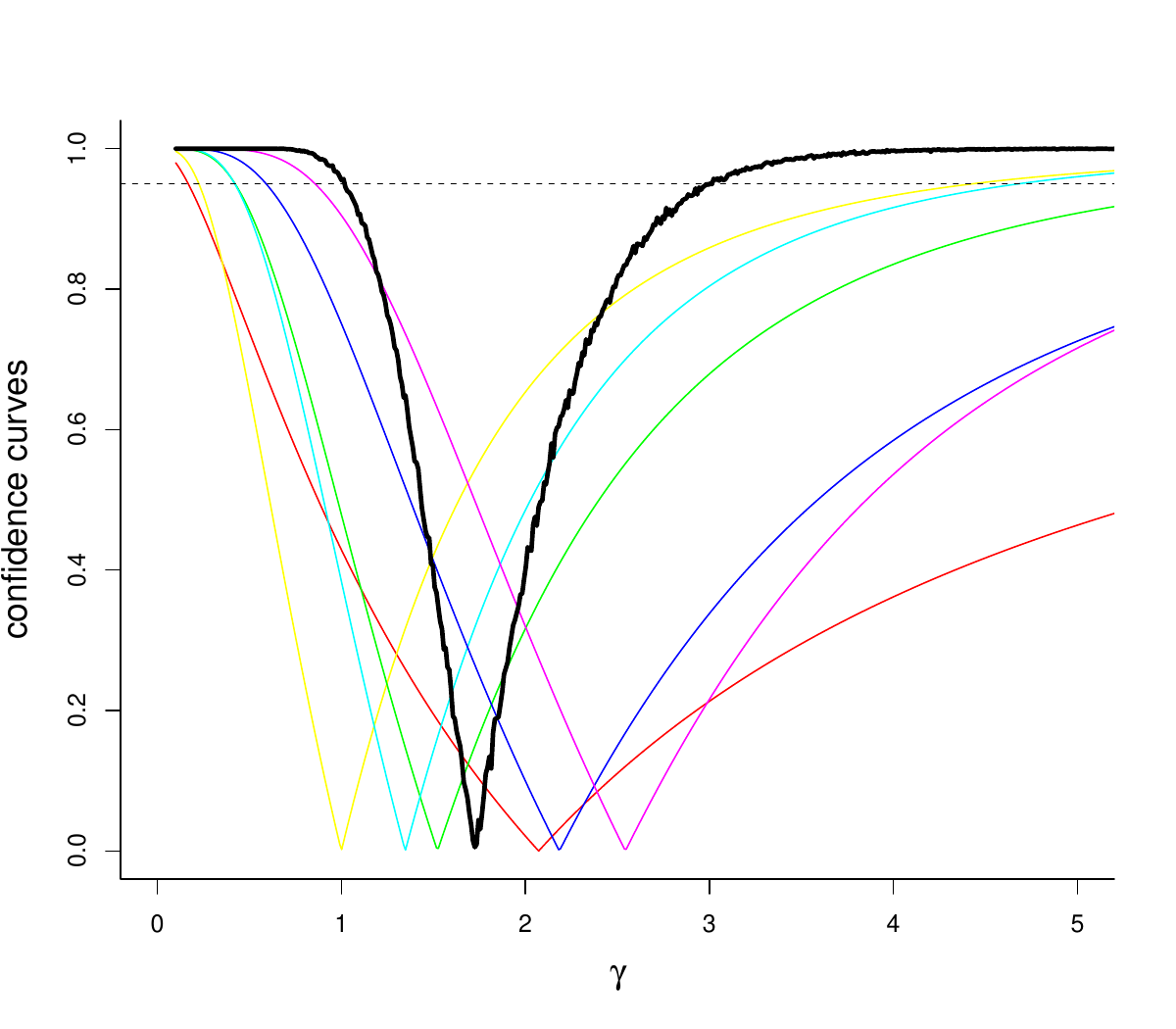}
\end{center}
\caption{Confidence curves for $\gamma$ for six Lidocaine studies (in colour) and combined confidence curve for all the studies (in black). A $95\%$
confidence interval for $\gamma$ can be read off (dashed line), [1.01,3.01]. }
\label{fig:1}
\end{figure}

Our modus of inference will be that of CDs.
A CD $C(\theta)$ for a parameter $\theta$ is a `distribution estimator', i.e.~a 
sample-dependent cumulative distribution function
over the parameter space summarising all aspects of frequentist inference for $\theta$ (for a thorough introduction to inference with CDs see \citet{clp2015} and \cite{xie2013}). 
In addition to presenting the results as a cumulative 
distribution function, it is fruitful to represent the distribution of confidence as a confidence curve, defined as 
\begin{equation*}
  \cc(\theta) = |1-2C(\theta)| .
\end{equation*}
The confidence curve is a funnel plot pointing to the median confidence estimate $\hat \theta = C^{-1}(\half)$;
also, confidence intervals for all levels $\alpha$ can be read off, see for example Figure \ref{fig:1}.

For parameters from models of the exponential class, there exist \textit{optimal} confidence distributions \citep[Ch.~5]{clp2015}. These are uniformly most powerful, 
in the sense of having lower expected  confidence loss than all other CDs for a large class of penalty functions and all parameter values
(see \cite{Schweder2002,clp2015}).
For both the binomial and Poisson model these optimal CDs are given in \citet[Ch.~14]{clp2015}. If the optimal (or other exact) CDs are hard or time-consuming to compute,
approximate CDs can be constructed. There are several methods for this, and we will present some examples in section 3. 
In section 2 the optimal CD for the fixed effect Poisson model will be provided, while the Poisson model with heterogeneous treatment effects (random effects) is treated in section 4. 
Section 5 offers some concluding remarks.
\smallskip

{\bf 2. Optimal confidence distribution for the fixed effect model}

The Poisson fixed effect model has the following log-likelihood, writing $z_i=y_{0,i}+y_{1,i}$, 
\begin{equation} \label{eq:1}
  \ell(\gamma,\lambda_1,\dots,\lambda_k) = \sum\limits_{i=1}^k \{y_{1,i}\log\gamma + z_i\log\lambda_i - (e_{0,i} + e_{1,i}\gamma)\lambda_i\} .
\end{equation}
This is a log-likelihood of the exponential family class. By theorems in \citet[Ch.~5]{clp2015} optimal inference for $\gamma$ is based on its sufficient statistic ($\sum Y_{1,i}$) given the 
ancillary statistics ($Z_1, \dots, Z_k$). The conditional distribution of $Y_{1,i} \midd Z_i$ is seen to be binomial,
\begin{equation} \label{eq:3}
  Y_{1,i}\midd (Z_i = z_i)  \sim \bin\left(z_i, \frac{e_{1,i}\gamma}{e_{0,i} + e_{1,i}\gamma}\right) .
\end{equation}
From this distribution we can construct CDs for $\gamma$ from each study (with half-correction for discreteness)
\begin{equation}  \label{eq:2}
  C_i(\gamma) =  P_{\gamma}(Y_{1,i} > Y_{1,i,\obs} \midd z_{i,\obs}) + \half P_{\gamma}(Y_{1,i} = Y_{1,i,\obs} \midd z_{i,\obs}) .
\end{equation}
Similarly, we can construct a combined CD for $\gamma$ for all the studies, which is uniformly most powerful according to \citet[Ch.~5]{clp2015},
\begin{equation*} \label{eq:4}
  C^{*}(\gamma) =  P_{\gamma}(B > B_{\obs} \midd z_{1,\obs},\dots,z_{k,\obs}) + \half P_{\gamma}(B = B_{\obs} \midd z_{1,\obs},\dots,z_{k,\obs}) .
\end{equation*}
where $B = \sum_{i=1}^k Y_{1,i}$, and $B_{\obs}$ is the observed value of $B$. This is a sum of binomially distributed variables with different probability parameters ($e_{1,i}\gamma/(e_{0,i} + e_{1,i}\gamma)$), 
and its distribution might easily be simulated.\\
As an illustration of this method, we have applied it to a medical datasets with six studies investigating death rates among heart attack patients \citep{normand1999}. The treatment group received the drug Lidocaine and the control group did not. 
The sample sizes and number of deaths in the six studies are listed in Table \ref{tab:1}. 

\begin{table}[h]
\tiny
\begin{center}
  \begin{tabular}{ccccc} \toprule
		Study & Sample size control & Events control & Sample size treatment & Events treatment\\ \midrule
    1 & 39 & 1 & 43 & 2 \\ 
    2 & 44 & 4 & 44 & 4 \\ 
    3 & 107 & 4 & 110 & 6 \\ 
    4 & 103 & 5 & 100 & 7 \\ 
    5 & 110 & 3 & 106 & 7 \\
    6 & 154 & 4 & 146 & 11 \\ \bottomrule
  \end{tabular}
\end{center}
  \caption{Sample sizes and number of deaths for the six Lidocaine studies.}
  \label{tab:1}
\end{table}

For the Lidocaine data, the combined CD for $\gamma$ is shown in Figure \ref{fig:1}, along  with CDs from each study. 
\vspace{-2em}
\smallskip

{\bf 3. Approximate confidence distributions for the fixed effects model}

\begin{figure}[h] 
\begin{center}
\includegraphics[scale=0.5]{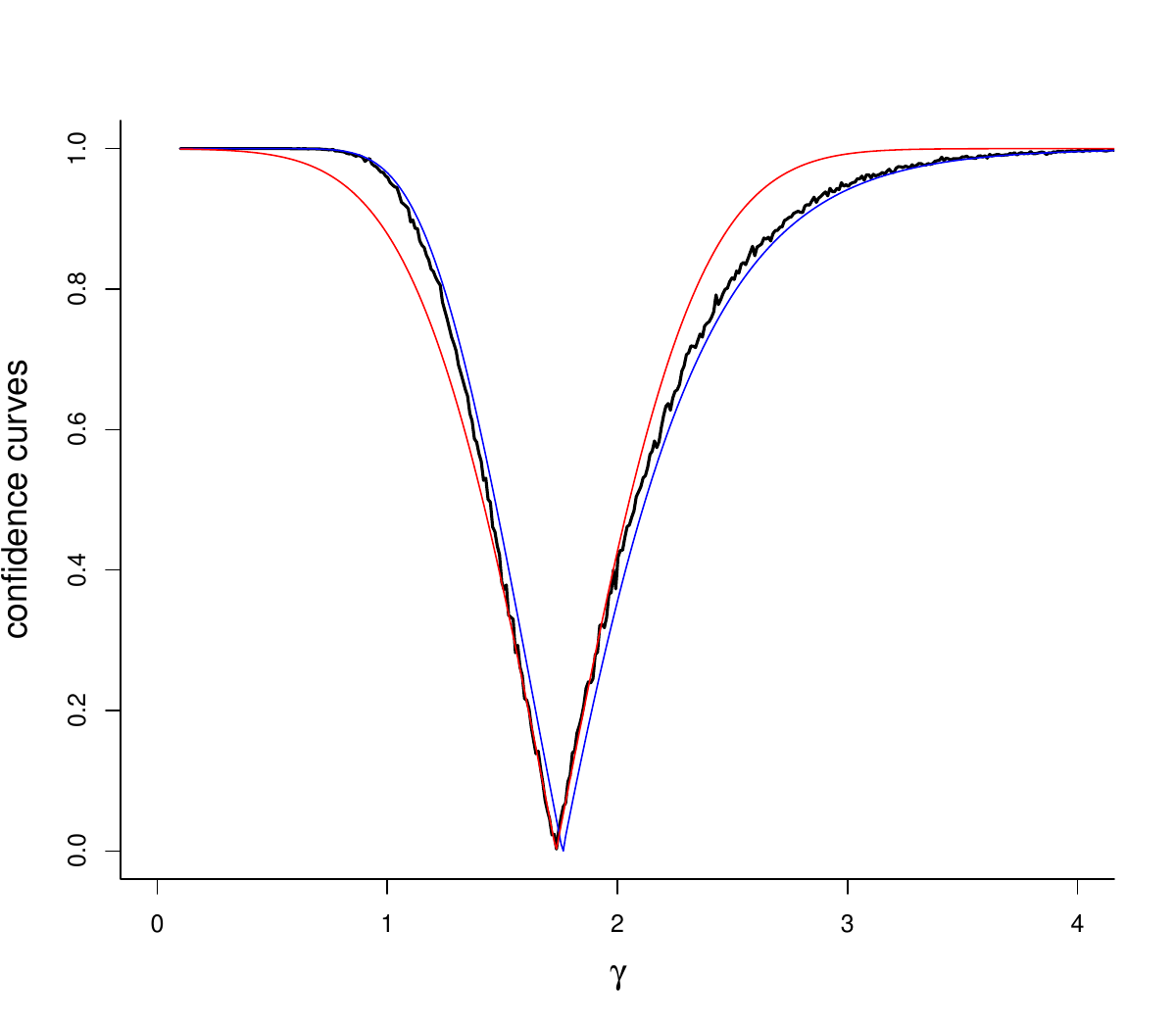}
\end{center}
\caption{Optimal confidence curve for $\gamma$ (in black) for the Lidocaine data, confidence curve based on the approximate distribution of the 
MCL (in red) and confidence curve based on profile deviance (in blue).}
\label{fig:2}
\end{figure}

The optimal CD can be slightly time-consuming to compute, but we can construct approximate CDs, offering a fast and often accurate alternative.
Consider the log-likelihood function of the conditional model following from \eqref{eq:3}. 
We learn that the maximum conditional likelihood (MCL) estimator is a monotone function of $B = \sum Y_{1,i}$. 
Since CDs are invariant with respect to monotone transformations, the optimal CD in the previous section is actually the same as the CD based on the MCL 
estimator $\hat{\gamma}_{\mcl}$. Ignoring half-corrections, we get 
\begin{equation*}
  C^{*}(\gamma) =  P_{\gamma}(B > B_{\obs} \midd z_{1,\obs},\dots,z_{k,\obs}) = P_{\gamma}(\hat{\gamma}_{\mcl} >  \hat{\gamma}_{\mcl,\obs}) .
\end{equation*}
For the Poisson model the MCL estimator of $\gamma$  is the solution of the following equation,
\begin{equation*}
  \sum\limits_{i=1}^k \frac{y_{1,i}(e_{0,i} + e_{1,i}\gamma) - z_ie_{1,i}\gamma}{\gamma(e_{0,i} + e_{1,i}\gamma)} =  0 .
\end{equation*}
Incidentally, the solution to the equation above is also the maximum likelihood (ML) estimator. Thus the unconditional and conditional likelihoods lead to the same estimators for this Poisson model.
This pans out differently for other models, as for example the model of binomial pairs \citep{breslow1981}. \\
Classical large samples methods for ML give us an approximate CD for $\gamma$, $C_{\appr}(\gamma)$. The distribution of 
$\hat{\gamma}_{\mcl}$ is approximately a normal ($\gamma, \hat{J}^{-1}$), with $\hat{J}$ the Hessian matrix associated with the log-likelihood function. This leads to the following approximate CD,
\begin{equation*}
   C_{\appr}(\gamma) = \Phi((\gamma - \hat{\gamma}_{\mcl})/\sqrt{\hat{J}^{-1}}).
\end{equation*}
Alternatively an approximate confidence distributions for $\gamma$ can be constructed by $\chi^2_1$-transforming the deviance function 
of the profiled likelihood (see \citet[Ch.~2-3]{clp2015}).
The $k+1$-dimensional log-likelihood function in \eqref{eq:1} is profiled by maximising over $\lambda_1, \dots, \lambda_k$, leading 
(in this case) to an explicit profile log-likelihood
\begin{equation*}
  \ell_{\prof}(\gamma) = \sum\limits_{i=1}^k \{y_{1,i}\log\gamma - z_i\log(e_{0,i} + e_{1,i}\gamma)\}.
\end{equation*}
The corresponding deviance function and the resulting confidence curve are 
\begin{equation*}
  D(\gamma) = 2\{\ell_{\prof}(\hat{\gamma}_{\ml}) - \ell_{\prof}(\gamma)\} \qquad \text{and} \qquad
  \cc_a(\gamma) = \Gamma_1(D(\gamma)) ,
\end{equation*}
where $\Gamma_1()$ is the cumulative distribution function of the $\chi^2_1$.
Both approximations work reasonably well, especially when there are many studies (large $k$), see Figure \ref{fig:2}.
\vspace{-2em}
\smallskip

{\bf 4. Optimal and approximate inference with heterogeneous treatment effects}

In meta-analysis of $2 \times 2$ tables, the  odds ratio from the binomial model is sometimes treated as a random effect (for example in the DerSimonian and Laird method).
Similarly, cell-counts from $2 \times 2$ tables can be modelled with a Poisson model with a non-constant $\gamma$ parameter. 
This allows the treatment effect $\gamma_i$ to vary, as in \eqref{eq:5}.
There are several possibilities regarding such models. 

\begin{figure}[h] 
\begin{center}
\includegraphics[scale=0.5]{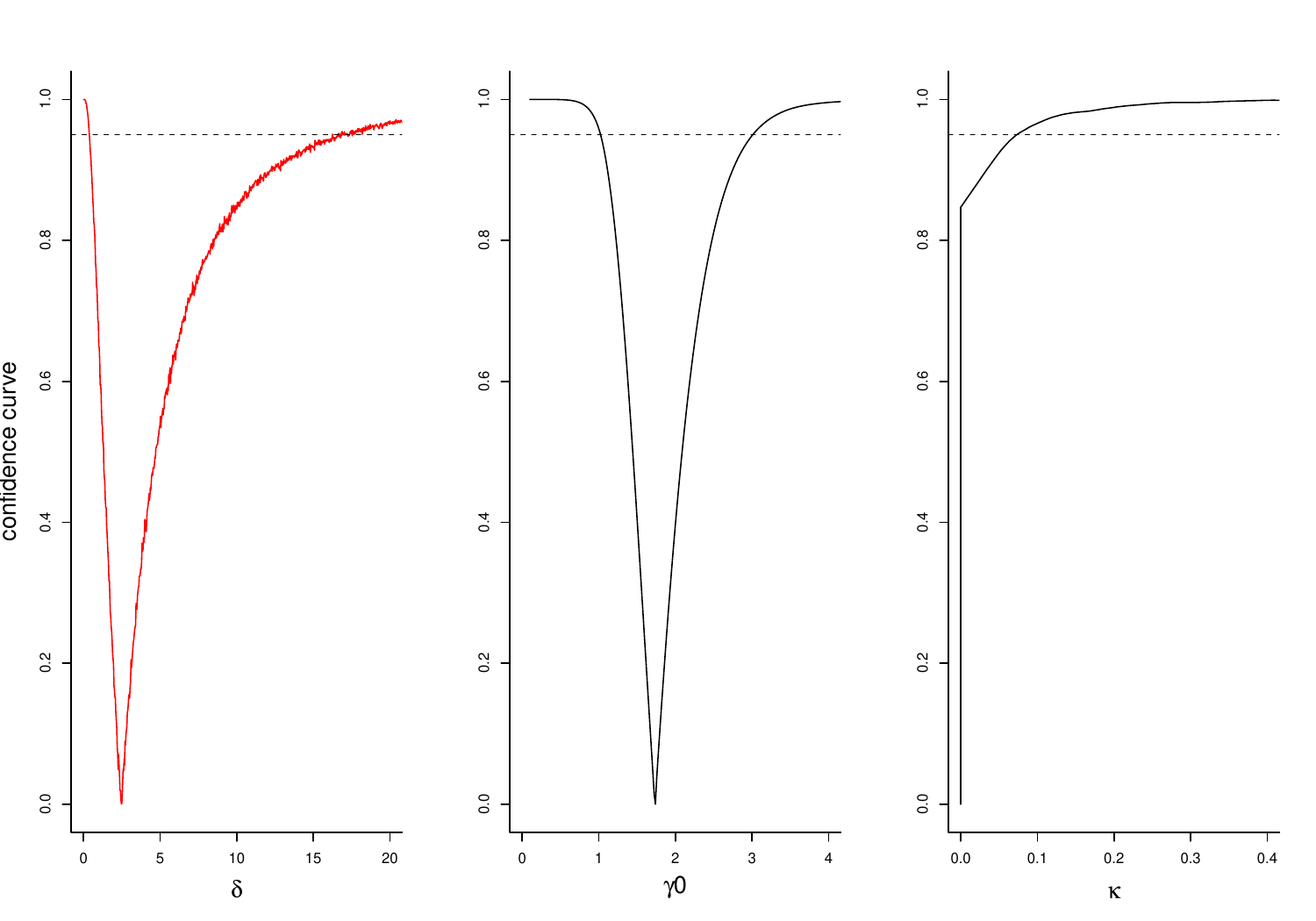}
\end{center}
\caption{Three confidence curves are shown. In panel 1, for $\delta$ between the two most different Lidocaine studies (in red); in panel 2 for $\gamma_0$; in panel 3 for $\kappa$.
The $95\%$ confidence intervals can be read off, $[0.37,17.81]$ for $\delta$, $[1.02,3.00]$ for $\gamma_0$ and $[0,0.072]$ for $\kappa$.
}
\label{fig:3}
\end{figure}

Our first suggestion is based on pairwise comparisons. Consider two studies at the time, for example study 1 and 2, with $Y_{0,1} \sim \pois(e_{0,1}\lambda_{1})$, $Y_{1,1} \sim \pois(e_{1,1}\lambda_{1}\gamma_1)$ 
and with $Y_{0,2} \sim \pois(e_{0,2}\lambda_{2})$, $Y_{1,2} \sim \pois(e_{1,2}\lambda_{2}\gamma_2)$. We can construct a CD for the ratio between the two treatment effects, $\delta=\gamma_2/\gamma_1$.
By arguments similar to those worked with in section 2, a power optimal CD for $\delta$ is found. 
It takes the form (with half-correction)
\begin{equation*}
  C_{1,2}(\delta) =  P_{\delta}(Y_{1,2}  > Y_{1,2,\obs} \midd w_{\obs},z_{1,\obs},z_{2,\obs}) + \half P_{\delta}(Y_{1,2}  = Y_{1,2,\obs} \midd w_{\obs},z_{1,\obs},z_{2,\obs}) ,
\end{equation*}
where $W=Y_{1,1}+Y_{1,2}$, $Z_1=Y_{0,1}+Y_{1,1}$ and $Z_2=Y_{0,2}+Y_{1,2}$. Thus we need the distribution of $Y_{1,2} \midd (W, Z_1, Z_2)$,
and we find that it is an eccentric hypergeometric distribution, which may be expressed as 
\begin{equation*}
  f(y_{1,2}  \midd (w, z_1, z_2) ) = \frac{\binom{z_1}{w-y_{1,2}}\binom{z_2}{y_{1,2}} \left(\frac{e_{0,1}e_{1,2}}{e_{1,1}e_{0,2}}\right)^{y_{1,2}}\delta^{y_{1,2}}}
  {\sum_{u=0}^{w} \binom{z_1}{w-u}\binom{z_2}{u} \left(\frac{e_{0,1}e_{1,2}}{e_{1,1}e_{0,2}}\right)^{u}\delta^{u}} 
  \text{ , for } y_{1,2} \text{ from $0$ to } \min(w,z_2).
\end{equation*}
In this way we can construct CDs for $\delta_{i,j}$ for all pairs among the six Lidocaine studies. In Figure \ref{fig:3} the CD for $\delta_{2,6}$ is displayed. 
These are the two studies with the most different median confidence estimates for $\gamma$ (see Figure \ref{fig:1}). The CD for this $\delta$ 
indicates that 1 is a likely value (the $95\%$ confidence interval is  $[0.37,17.81]$).
Hence the $\gamma$s in the Lidocaine data do not seem sufficiently different to justify the use of a random effects model. 
\vspace{-2em}

The variation among the treatment effects from different studies can also be investigated in other ways. A natural suggestion is to model the variation of the $\gamma_i$ 
with a beta distribution for the probabilities $\pi_i$ from the conditional distribution of $Y_{1,i} \midd (Z_i = z_i)$, 
\begin{equation*}
  Y_{1,i} \midd (Z_i = z_i, \gamma_i)  \sim \bin(z_i, \pi_i) \quad \text{with} \quad  \pi_i = \frac{e_{1,i}\gamma_i}{e_{0,i} + e_{1,i}\gamma_i} ,
\end{equation*}
where in addition $\pi_i \sim \beta\{\tau\pi_{0,i}, \tau(1-\pi_{0,i})\}$ and $\pi_{0,i}=e_{1,i}\gamma_0/(e_{0,i} + e_{1,i}\gamma_0)$. Letting $\kappa = (\tau + 1)^{-1}$,
we have 
\begin{equation*}
    \E(\pi_i) = \pi_{0,i} \qquad \text{and} \qquad \Var(\pi_i) = \kappa\pi_{0,i}(1-\pi_{0,i}) .
\end{equation*}
If $\tau = \infty$, then $\kappa$ is equal to zero and we are back in the Poisson model with a common $\gamma$.
We can construct approximate confidence curves for $\gamma_0$ by using profile deviance methods as described in section 3, 
\begin{equation*}
  \cc(\gamma_0) = \Gamma_1(D_1(\gamma_0)), \qquad D_1(\gamma_0) = 2\{\ell_{1,\prof}(\hat{\gamma}_0) - \ell_{1,\prof}(\gamma_0)\}.
\end{equation*}
For $\kappa$, $\cc(\kappa) = | 1- 2 P_{\kappa}(Q_{\min} \ge Q_{\min,\obs})|$ is a valid confidence curve, since it is (by construction) nearly uniformly distributed at the true parameter value.
We use the natural statistic
\begin{equation*}
  Q_{\min} = \min_{\gamma_0} \sum\limits_{i=1}^k \frac{\{y_{1,i}-z_if_i(\gamma_0)\}^2}{z_if_i(\gamma_0)\{1-f_i(\gamma_0)\}} \text{, where} \quad f_i(\gamma_0) = \frac{e_{1,i}\gamma_0}{e_{0,i} + e_{1,i}\gamma_0},
\end{equation*}
and calculate it for the observed data (finding $Q_{\min,\obs}$) and for a number of $y_{1,i}$ vectors, simulated under different $\kappa$ values, with fixed $z_i$ and $\hat \gamma_0$.\\
The CD for $\gamma_0$ allows us to make inference about the increase in risk related to the the treatment in question. 
While the CD for $\kappa$ is used to assess the variation in the treatment effect. If the estimate for $\kappa$ is small and the CD for $\kappa$
has a significant point mass at zero, then the variation among the treatment effect is small and 
we can choose to use the fixed effect model from section 2. Otherwise, the random effect model may be preferable. For the Lidocaine example, the CD estimate for $\kappa$ is zero and the point mass at
zero is 0.85 (Figure \ref{fig:3}). 
This supports our conclusion for the $\delta$ analysis above, that there is little variation among the $\gamma$s for the Lidocaine studies, 
and that the fixed effect model is the most appropriate.
\smallskip

{\bf 5. Concluding remarks}\\
There are many other aspects of meta-analysis of $2 \times 2$ tables (in the CD framework) worth considering. First,
we have applied our CD-methods to several other datasets, including a small dataset from \cite{narum2014}, 
with five studies investigating whether drugs of a certain type (corticosteroids) are related to 
gastrointestinal bleeding for patients in ambulatory care. These data poses some challenges for many traditional meta-analysis methods because most of the studies have zero event entries, either in the treatment group
or the control group. Many existing methods demand 0.5 corrections to zero events in such situations, 
but these kinds of corrections can give large biases and are not recommended \citep{bradburn2007}.
One of the strengths of our method  is that it can produce a combined confidence distribution in this case, without needing to add anything to the data.
Secondly, we have implemented the alternative CD based combination method from \cite{liu2014}, and compared it to our methods. The two classes of methods
can yield similar results, but ours yield stronger confidence. Lastly we have considered extensions to $2 \times 3$ tables (which the Poisson model can handle without problems)
and alternative ways to model the heterogeneous treatment effects (for example with a Poisson-gamma model). \\

{\bf References}

\bibliographystyle{apalike2} 
\bibliography{ref_rio}

\end{document}